\documentclass{emulateapj}

\usepackage{natbib}

\newcommand{\myemail}{ebuenzli@email.arizona.edu}

\slugcomment{Accepted for publication in ApJL, October 24 2012}

\shorttitle{Vertical Atmospheric Structure in a Variable Brown Dwarf}

\shortauthors{Buenzli et al.}

\begin{document}

\title{Vertical Atmospheric Structure in a Variable Brown Dwarf: Pressure-dependent Phase Shifts in Simultaneous {\it Hubble Space Telescope$-$Spitzer} Light Curves} 

\author{Esther Buenzli$^1$, D\'{a}niel Apai$^{1,2}$, Caroline V. Morley$^{3}$, Davin Flateau$^{1,2}$, Adam P. Showman$^2$, Adam Burrows$^{4}$, Mark S. Marley$^5$, Nikole K. Lewis$^{2,6,}$\footnotemark[8], and I. Neill Reid$^{7}$} 

\footnotetext[8]{Sagan Fellow}

\affil{$^1$Department of Astronomy and Steward Observatory, University of Arizona, Tucson, AZ 85721, USA; \myemail}
\affil{$^2$Department of Planetary Sciences and Lunar and Planetary Laboratory, University of Arizona, Tucson, AZ 85721, USA}
\affil{$^3$Department of Astronomy and Astrophysics, University of California, Santa Cruz, Santa Cruz ,CA 95060, USA}
\affil{$^4$Department of Astrophysical Sciences, Princeton University, Princeton, NJ 08544, USA}
\affil{$^5$NASA Ames Research Center, Moffett Field, CA 94035, USA}
\affil{$^6$Department of Earth, Atmospheric and Planetary Sciences, Massachusetts Institute of Technology, Cambridge, MA 02139, USA}
\affil{$^7$Space Telescope Science Institute, Baltimore, MD 21218, USA}

\begin{abstract}
Heterogeneous clouds or temperature perturbations in rotating brown dwarfs produce variability in the observed flux. We report time-resolved simultaneous observations of the variable T6.5 brown dwarf 2MASSJ22282889$-$431026 over the wavelength ranges $1.1-1.7$ $\mu$m and broadband 4.5 $\mu$m. Spectroscopic observations were taken with Wide Field Camera 3 on board the {\it Hubble Space Telescope} and photometry with the {\it Spitzer Space Telescope}. The object shows sinusoidal infrared variability with a period of 1.4 hr at most wavelengths with peak-to-peak amplitudes between 1.45\% and 5.3\% of the mean flux. While the light curve shapes are similar at all wavelengths, their phases differ from wavelength to wavelength with a maximum difference of more than half of a rotational period. We compare the spectra with atmospheric models of different cloud prescriptions, from which we determine the pressure levels probed at different wavelengths. We find that the phase lag increases with decreasing pressure level, or higher altitude. We discuss a number of plausible scenarios that could cause this trend of light curve phase with probed pressure level. These observations are the first to probe heterogeneity in an ultracool atmosphere in both horizontal and vertical directions, and thus are an ideal test case for realistic three dimensional simulations of the atmospheric structure with clouds in brown dwarfs and extrasolar planets. 
\end{abstract}

\keywords{brown dwarfs --- stars: atmospheres ---  stars: individual (2MASSJ22282889$-$4310262) --- stars: variables: general}

\section{Introduction}

\setcounter{footnote}{8}

Flux variability in brown dwarfs has been attributed to heterogeneous cloud cover that modulates the light curve through rotation. Two early T dwarfs have recently been discovered as photometrically variable at the multiple percent level \citep{artigau09,radigan12}. These objects lie at the L/T transition between cloudy L dwarfs and mostly clear late T dwarfs, where clouds have dissipated or sunk below the photosphere. The mechanism for cloud dispersion is not well understood, and different models have been proposed that invoke cloud thinning and rain-out \citep[e.g.,][]{knapp04, tsuji05, burrows06, saumon08} or patchy clouds \citep{marley10}. Variability indicates heterogeneous cloud structure, but models with cloud holes that allow flux from hot deep regions to escape do not fit the measured time-resolved spectroscopy in early T-dwarfs \citep{apai12}. A mixture of thin and thick clouds provides a better match.   

Variability is not limited to L/T transition dwarfs, though these appear to show the highest variability level. $I$ band surveys found several variable L dwarfs \citep[e.g.][]{bailerjones01, gelino02}, and \citet{clarke08} reported variability for two T dwarfs beyond T5. While later T dwarf spectra have been modeled as clear atmospheres with reasonable success \citep[e.g.][]{allard03,stephens09}, color dispersion and variability indicate that clouds, such as sulfide clouds \citep{morley12}, may persist in these atmospheres. 

The T6.5 dwarf 2MASSJ22282889-4310262 (hereafter 2M2228) was discovered by \citet{burgasser03}. \citet{clarke08} measured periodic variability, likely due to rotation with a period of 1.43~hr and a peak-to-valley amplitude of $1.4\%\pm0.14$\% of the mean flux in $J$ band. Its parallax was measured as $94\pm7$~mas by \citet{faherty12}, resulting in a distance $d=10.64\pm0.79$~pc. 

Here we report time resolved high-precision observations of 2M2228: simultaneous {\it HST}/WFC3 spectroscopy and {\it Spitzer}/IRAC photometry that for the first time simultaneously probe variability at different altitudes in the atmosphere of a brown dwarf. The observations reveal a phase shift between wavelengths that is evidence for vertical heterogeneity in the atmosphere. They provide unique constraints on the heterogeneous cloud or temperature structure in both the horizontal and vertical directions. 

\section{Observations}

We observed 2M2228 on 2011 July~07 with the infrared channel of the Wide Field Camera 3 (WFC3) on the {\it Hubble Space Telescope} ({\it HST}) and the Infrared Array Camera (IRAC) of the {\it Spitzer Space Telescope} (GO-12314, GO-70170; PI: D. Apai). {\it HST} observed between BMJD(UTC) 55749.528 to 55749.886, {\it Spitzer} between 55749.766 and 55750.002, providing an overlap of almost three hours.

With {\it HST}, we obtained spectral time series with the G141 grism for the wavelength range $1.05-1.7$ $\mu$m over six consecutive {\it HST} orbits, spanning 8.58~hr. For each orbit of 96 minutes, we observed the target for 35.6 minutes, obtaining nine exposures of 224~s. For the rest of the time, the target was occulted by Earth.

The infrared channel of WFC3 is a Teledyne HgCdTe detector with 1024x1024 pixels. We used the 256 x 256 pixels subarray with a field of view of approximately 30 x 30\arcsec to avoid WFC3 buffer dump overheads. Each exposure is constructed from multiple non-destructive read-outs. We used the SPARS25 readout mode with 11 reads per exposure. Each sub-readout has an exposure time of 22.34 s after an initial first short read at 0.27~s. The maximum number of counts in a pixel at the end of an exposure was 27,000 counts, well below the detector half-well capacity. 

The first-order spectrum has a width of  $\approx140$~pixels and a dispersion of 4.65~nm~pixel$^{-1}$. The zeroth and second orders were not recorded. A direct image was taken for wavelength calibration through the F132N narrow band filter at the beginning of each orbit. We kept the spectra on the same pixels for all orbits to avoid systematic errors from imperfectly corrected pixel-to-pixel sensitivity variations. 

The {\it Spitzer} observations were taken with IRAC channel 2 at 4.5 $\mu$m (warm mission phase). Channel 2 has a 256 x 256 pixel InSb detector with a field of view of 5.2 x 5.2~arcmin. We acquired 2876 individual measurements with 4.4~s exposure time and cadence of 7.2~s, covering 5.75~hr. 

\section{Data reduction}

\subsection{{\it HST}/WFC3}

We based our data reduction on the combination of standard WFC3 pipeline, the PyRAF software package aXe\footnote{http://axe.stsci.edu/axesim/}, and custom IDL routines. We summarize here the most important points, additional details can be found in \citet{apai12}.

The WFC3 pipeline \texttt{calfw3} produces flux calibrated, two-dimensional spectral images with flagged bad pixels. We start our reduction from the \texttt{.ima} files which contain an array of each subread for each exposure. We discard the short zero read, resulting in 10 images of 22.34~s per exposure. After the full reduction and spectral extraction, we average the 10 spectra obtained per exposure into one to reduce the noise. 

We correct bad pixels by interpolating over neighbors and cosmic rays by replacing outliers in a subexposure sequence with the median pixel value over the sequence. For the background subtraction, flat fielding and extraction of the two dimensional spectra, we use aXe,$^9$ developed specifically for {\it HST} slitless spectroscopy. We choose the spectral extraction width to be three times the FWHM of the vertical spectral profile. The FWHM is determined individually for each orbit by averaging all spectra and collapsing the result into a one-dimensional (1D) vertical profile, to which we fit a Gaussian. The mean extraction width was $5.651\pm0.013$~pixels. The positional differences of the spectra over a visit were determined with cross-correlation to be smaller than 0.02~pixels.   

The extracted 1D spectra are flux calibrated with the G141 sensitivity curve, which has an estimated uncertainty of 1\% \citep{kuntschner11}. We calculate the flux uncertainty from the read-out noise and the photon noise of the source and background. The time of observation is calculated as the mid-time between sub-reads from the header keywords ``ROUTTIME''  and ``EXPTIME''. The times are converted from Modified Julian Date (MJD) to Barycentric Modified Julian Date (BMJD-UTC) with the barycen.pro IDL routine to allow comparison with the {\it Spitzer} data, and double-checked with the utc2bjd.pro routine. These routines are accurate to well below a second \citep{eastman10}. The spectra have a useable wavelength range of $1.05-1.7$~$\mu$m and a resolution $R=130$. 

We identified and corrected two detector systematics at sub-percent levels. First, in each batch of sub-reads of an exposure the measured flux dropped by $\approx1\%-2\%$ from the first to last su-bread for the bright pixels in a spectrum, likely a result of charge diffusion. A second effect was an exponential ramp at the beginning of each orbit, which was found to be independent of count rate and wavelength. Both effects were corrected with analytic functions \citep{apai12} derived from a non-variable source (ramp) or an average over all sources (flux drop). 

\subsection{{\it Spitzer}/IRAC}

The raw {\it Spitzer} data were processed with the IRAC pipeline ver. 19.0.0, which performs dark subtraction, flat-fielding, flagging of bad pixels and flux calibration. We masked columns that were significantly noisier than the rest of the pixels, and applied a position-dependent flux correction. We determined the source center with the gcntr.pro routine with an FWHM of 4 pixels. We performed aperture photometry with the aper.pro routine with an aperture of 2 pixels radius, and the sky annulus between 12 and 20 pixels radius, which minimized the noise. We identified and removed frames with cosmic ray hits across the source by checking for anomalous centroid positions and flux outliers of more than 5$\sigma$. To correct for intrapixel sensitivity variations, we used a pixel mapping method \citep{knutson12,lewis12}. We determined the noise pixel parameter $\tilde{\beta}$ for each exposure and created a pixel flux map using each measurement's 50 nearest spatial neighbors. After applying the flux correction, we binned the data into 5 minute bins to reduce the noise. We calculated the error from the shot noise and the statistical uncertainty of the sky subtraction. The remaining systematic errors after the intrapixel sensitivity correction were smaller than the random errors in our binned data. The time of observations in BMJD(UTC) is given in the {\it Spitzer} data header \mbox{BMJD\_OBS}, and was adjusted to correspond to the mid-exposure time.  
 
\section{Results}

\begin{figure*}
\epsscale{1.1}
\plotone{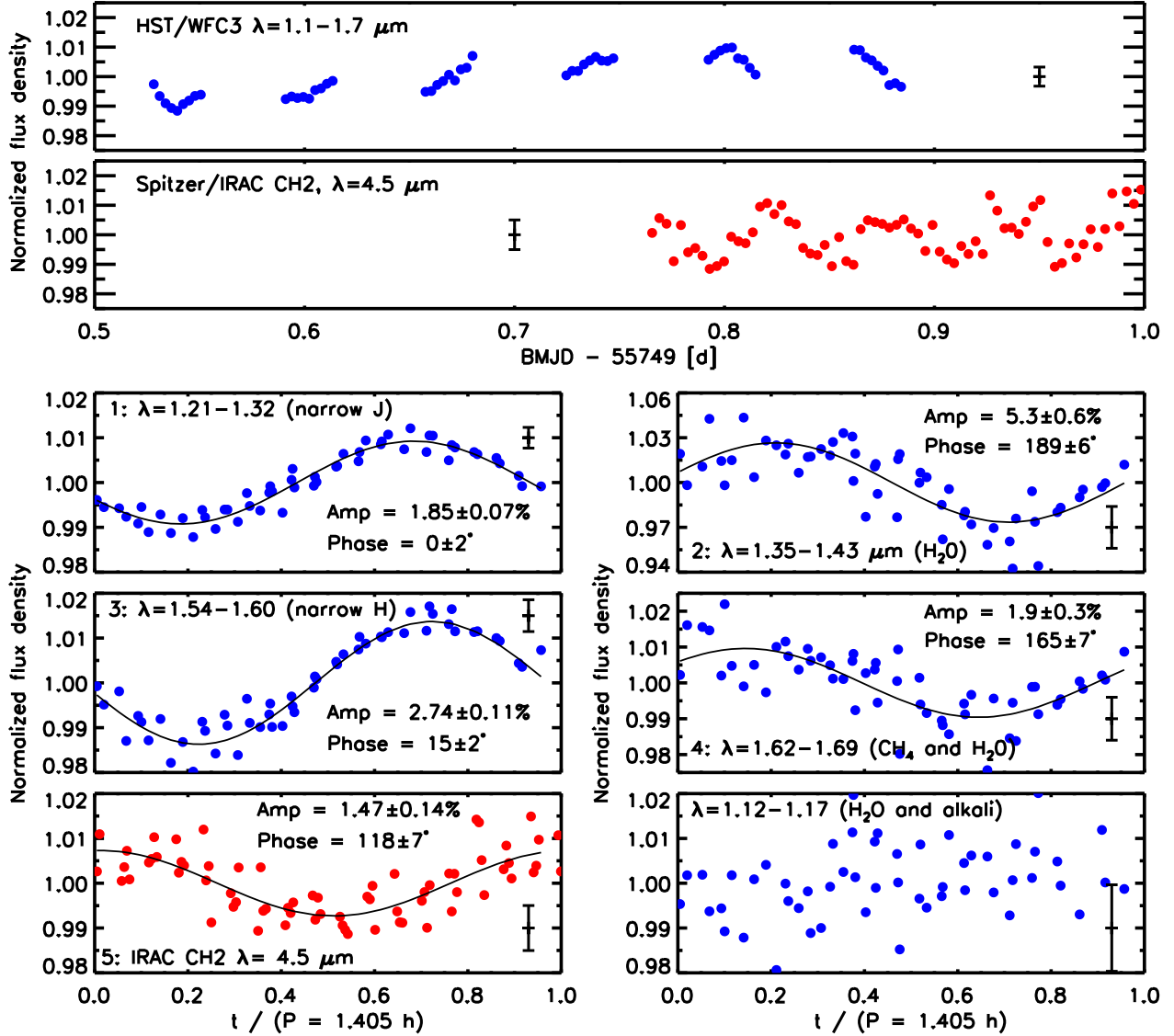}
\caption{{\it HST} (blue) and {\it Spitzer} (red) observations of 2M2228 integrated over different bandpasses. Top: {\it HST} time series integrated over $1.1-1.7$~$\mu$m and {\it Spitzer} photometry at 4.5~$\mu$m as a function of barycentric time. Bottom: period folded light curves in six wavelength regions folded with the rotation period $P=1.405$ hr. A representative error bar at each wavelength is a combination of random and systematic errors. The black curve is the best sine fit. The phase is given with respect to the phase of the narrow $J$ light curve.\label{fig:allres}}
\end{figure*}

2M2228 shows periodic variability in both the {\it HST} and {\it Spitzer} data sets. Figure~\ref{fig:allres} (upper panels) shows the {\it HST} time series integrated over the full wavelength range compared to the {\it Spitzer} photometry. 2M2228 is a fast rotator with a rotation period of only $1.405\pm0.005$ hr determined from a sine fit to the {\it HST} data. Each {\it HST} orbit of 96 minute samples part of a subsequent rotation period with a shift by 1/6th of a period. Six orbits cover the full rotation phase, and 90\% of the phases are sampled at least twice. The {\it Spitzer} observations continuously cover four adjacent rotation periods, two of which partially overlap in time with the {\it HST} data. The derived period is $1.38\pm0.03$~hr. We adopt the more precise {\it HST} period for further analysis. 

\begin{figure*}
\epsscale{1.1}
\plotone{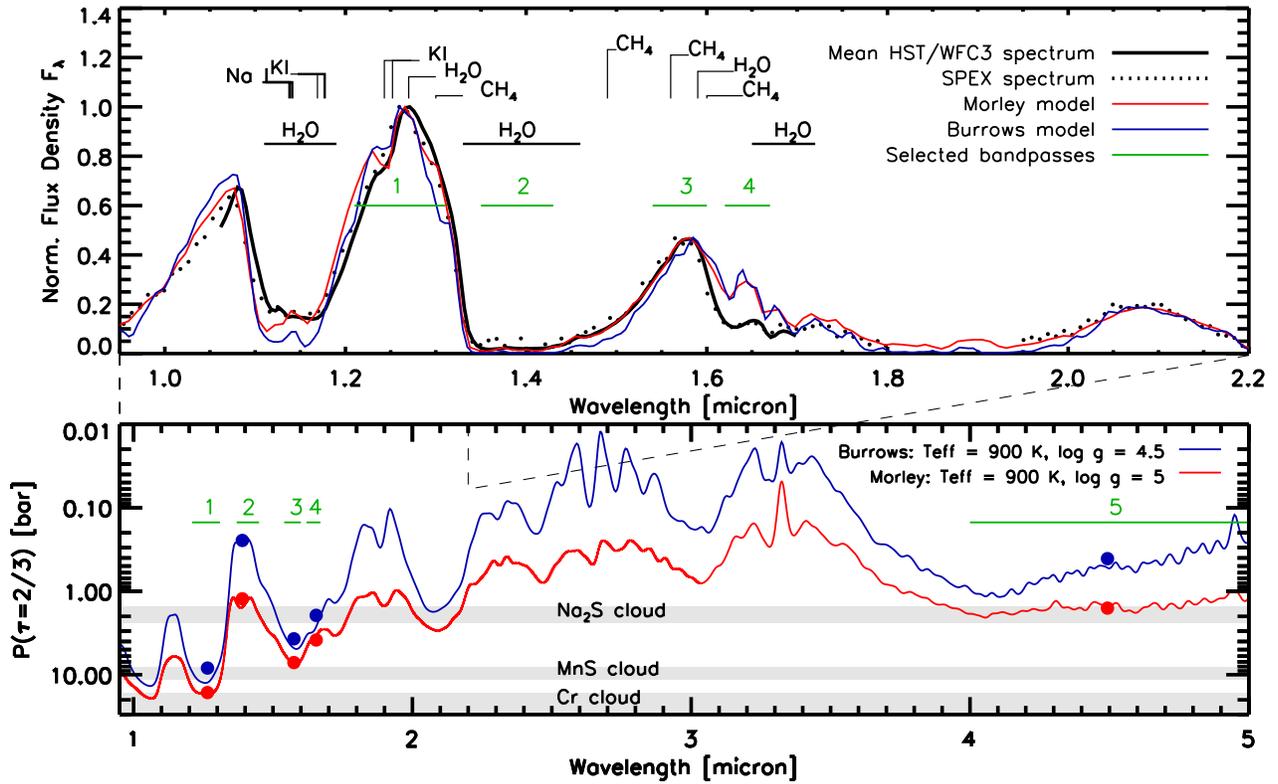}
\caption{Top: normalized mean {\it HST}/WFC3 spectrum of 2M2228 together with the SPEX spectrum of the source by \citet{burgasser04}, and best-fit atmospheric models by Morley and Burrows (see text). Bottom: pressure level probed at optical depth $\tau=2/3$ as a function of wavelength for the two models. Cloud layers are indicated for the Morley model. Selected bandpasses to study the variability are indicated by the horizontal green lines (cf. Figures \ref{fig:allres} and \ref{fig:phaseshift}), and averaged pressures in these bandpasses are marked with dots.\label{fig:specs}}
\end{figure*}

The mean {\it HST} spectrum is plotted in Figure \ref{fig:specs} (top panel) together with a ground-based SPEX spectrum \citep{burgasser04} and compared to models (see Section \ref{sect:models}). The spectra are dominated by deep water absorption features with higher flux in the $J$ and $H$ band windows. Other features include CH$_4$ and K I and Na I. We study the variability as a function of wavelength by integrating the spectra for a number of characteristic regions. The bandpasses are selected such that the flux is emerging from a well-defined pressure region in the atmosphere. Integrated and period-folded light curves at these wavelengths, together with the {\it Spitzer} photometry at 4.5~$\mu$m, are shown in the lower panels of Figure \ref{fig:allres}.

\begin{figure*}
\epsscale{1.15}
\plotone{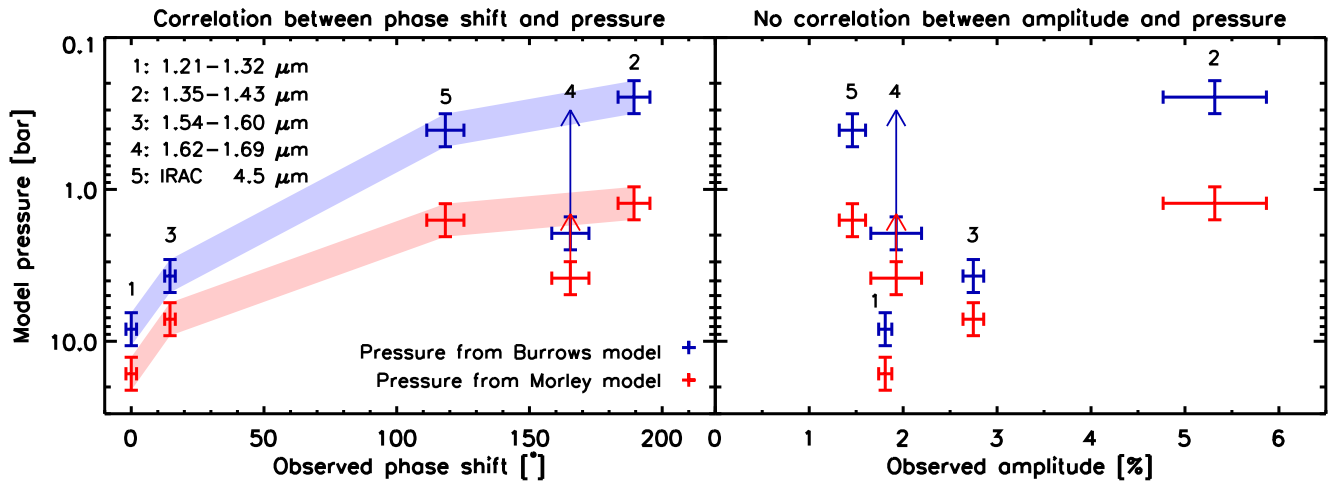}
\caption{Phase shifts (left) and amplitudes (right) of the observed 2M2228 light curves as a function of the dominantly probed pressure. There is a correlation between phase shift and pressure (shaded regions), but not between amplitude and pressure. Phase shifts in degrees are relative to the $J$ band light curve. The pressure in each wavelength region is derived from the two models presented in Figure~\ref{fig:specs}. The assumed width of the pressure region is half a scale height, the horizontal width of the bar is the $\pm1\sigma$ observational error. The pressure for bandpass~4 is likely overestimated due to underestimated methane opacity in the model and should be at lower pressure (arrow), making it consistent with the trend of the phase shift.\label{fig:phaseshift}}
\end{figure*}

The light curve shapes in all wavelength regions are well-fit by sinusoidal functions of the form $F(t)=1+A_0\cos{t}+B_0\sin{t}$. We determine amplitudes, phases and errors from sine fits by bootstrapping (Figures \ref{fig:allres} and \ref{fig:phaseshift}). We do not find significant variability for the wavelength range $1.12-1.17$. Because of high noise due to low grism sensitivity and poor correction of systematics there may be undetected variability on the 1\%-2\% level. We do not include this wavelength range in our further analysis. The peak-to-valley amplitude levels range between $\approx$1.5\% in the IRAC band more than 5\% in the deep water band around 1.4~$\mu$m. 

Most interestingly, we find significant shifts in the phase of the light curve as a function of wavelength. This is evident in the lower panels of Figure \ref{fig:allres}. All light curves are period-folded with reference to the same time $t_0$, taking the narrow J band light curve as zero reference for the phase. The narrow $H$ band curve shows a small delay in phase of $14^\circ.6\pm2^\circ$, or $3.4\pm0.5$ minutes. The 4.5~$\mu$m phase curve is delayed much more: $118\pm7^\circ$ or $27.7\pm1.6$ minutes. The deep absorption bands show the largest phase difference to $J$: $165^\circ\pm7^\circ$ for $1.62-1.69$ $\mu$m and $189^\circ\pm6^\circ$ for $1.35-1.43$ $\mu$m, more than half a rotation period.

\section{Modeling}
\label{sect:models}

Although current state-of-the-art atmospheric models of brown dwarfs are limited to a 1D treatment of radiative transfer and convection, and do not include local heterogeneities, they can provide globally averaged atmospheric quantities. We fit 1D models to the mean spectrum, and to the SPEX spectrum from \citet{burgasser04} for broader coverage into the $Y$ and $K$ bands, to derive basic atmospheric parameters and determine the pressure levels that are probed by different wavelengths. We use two independent sets of T-dwarf models to test the robustness of our conclusions. The first are models by \citet{morley12}, based on the \citet{ackerman01} cloud models and including sulfide and chromium clouds. The models use a sedimentation efficiency parameter $f_{sed}$, where larger $f_{sed}$ corresponds to larger grains and thinner clouds.  The second are models by \citet{burrows06} and adaptions thereof from \citet{hubeny07}, \citet{currie11}, and \citet{madhu11} that include different chemical compositions, cloud thickness prescriptions and grain sizes. 

We find the best spectral match using a Morley model (Figure~\ref{fig:specs}) with effective temperature $T_{eff}=900$ K , surface gravity of $\log g=5$, and $f_{sed}=5$, though $\log g=4.5$ and $f_{sed}=3$ (thicker clouds) provide an almost equally good fit. This model includes thin Na$_2$S, MnS, and Cr clouds. The overall spectral fit is good, with the main discrepancy in the $H$ band around 1.65~$\mu$m. This discrepancy is common in many sources \citep{morley12}, and is likely the result of incomplete methane opacities. A second discrepancy are alkali absorption features at 1.24~$\mu$m and 1.12~$\mu$m, whose depths are overestimated by all models. 

The best Burrows model is an ``iron" cloud model with $T_{eff}=900$ K and $\log g=4.5$, where iron provides representative optical properties. It uses the ``Model A" cloud prescription \citep{burrows06}, where the cloud density is exponentially declining towards the top. The overall spectral shape is reproduced reasonably well, although with similar and somewhat larger discrepancies than the Morley model. Again, other models with different cloud prescriptions or chemical abundances provide similar fits. We find that the main conclusions of this work do not depend on the detailed cloud model structure. 

From our best-fit models we infer the pressure levels that are typically probed at different wavelengths. With this information we translate the phase shift in the light curves as a function of wavelength to a phase shift as a function of pressure. Together with the longitudinal information from the light curve variability, our observations constrain heterogeneity both in the horizontal and vertical direction. 

Figure \ref{fig:specs} (bottom) shows the pressure levels at optical depth $\tau=2/3$, with cloud levels for the Morley model indicated. In reality, a range of pressures is probed simultaneously, and can be described by contribution functions \citep[see, e.g.,][]{marley12}. Instead of using these functions, because we are only looking for trends as a function of probed pressure region, for each wavelength band we derive an average pressure level and assign half an atmospheric scale height for the width of the region. 

In Figure \ref{fig:phaseshift} (left), the phase shift is shown as a function of pressure for both models. We find that the phase delay increases strongly with decreasing pressure (higher altitude) for the four bandpasses where the model fit is good. For the 1.62-1.69 $\mu$m bandpass, the model fit is poor because the methane opacities are likely underestimated. This means that the actual probed pressure is lower (higher altitude) than determined from the model, making that point consistent with the trend. Qualitatively, the trend is the same for both models, only shifted to lower pressures for the Burrows model, mostly because of the lower gravity of the model. 

The right panel of Figure \ref{fig:phaseshift} gives the peak-to-peak amplitude of the variability as a function of pressure. There is no obvious correlation between amplitude and probed pressure level. 

\section{Discussion}

We have discovered sinusoidal infrared variability in a late T-dwarf with phase offsets between different wavelengths. From comparing atmosphere models to the spectrum, we find a clear trend of the phase shift with the pressure probed at respective wavelengths: the lower the pressure of the atmospheric layer, the larger the phase lag with respect to the highest pressure layer, up to half a rotation period. We consider three cases that may cause infrared variability to be out of phase: (I) change in the temperature-pressure profile $T(p)$ without change in opacities, (II) change in the opacities (clouds and/or gas) without change in $T(p)$, or (III) an intermediate scenario. 

{\it Case (I):} Define $\Delta T$ as the horizontal temperature difference on isobars. To get the observed variability requires that regions where $\Delta T$ is positive aloft exhibit negative $\Delta T$ at depth, and vice versa. Hotter regions may result from subsidence, which advects high-entropy gas from above, while cooler regions may
result from ascent, which advects low-entropy gas from below.  This suggests a ``stacked cell'' scenario \citep[e.g.][]{fletcher11} where, on one side of the brown dwarf, subsidence at depth accompanies ascending motion aloft---and vice versa on the other side.

{\it Case (II):} In this scenario, changes to the brightness temperature result from the $\tau=2/3$ level moving up or down. It seems unlikely that opacity changes due to clouds alone can produce the observed trend for the phases of the light curves. The strongest variability, ~5\%, is seen at the lowest pressure level, where no significant cloud opacity is expected from the models. If present, cloud opacity variations at such high altitude would affect the emission in other bands and cause the variability to be mostly in phase. A second, lower level cloud deck would need to show much stronger spatial variations in opacity of opposite phase to compensate for the opacity variations of the high cloud layer.

Gas opacity is strongly wavelength dependent. In principle, regions with increased gas opacity at 1.4 $\mu$m could have decreased opacity
in $J$ and $H$, resulting in a phase difference in the light curves between these wavelengths. Still, pure gas opacity changes are
unlikely since clouds likely still play a role in these atmospheres as found from the model comparison. A mixed scenario could be plausible, where
the variability at higher pressure regions stems from changes in cloud opacity, while at low pressure the gas opacity varies with different
phase. It is not clear which gas could produce such opacity changes. Water does not condense at the given temperatures, and CO and CH$_4$ should largely be quenched \citep{hubeny07}, making spatial variations difficult. A condensing species, e.g. those considered in the \cite{morley12} models, where condensation induces spatial variations in the gas abundance, is a possibility.

{\it Case (III):} Of various possibilities, a plausible scenario involves a heterogeneous deep cloud controlling the variability at $J$ and $H$, with horizontal variations in temperature aloft producing the variations in bands sensing high altitudes. Again, a ``stacked cell'' scenario as in case (I) would be a possible cause, where a region of
subsidence that affects the deep cloud is overlain by a region of ascent that affects the high-altitude temperature. Such a stacked-cell scenario has been suggested for the tropospheres of Jupiter \citep{ingersoll00} and Saturn \citep{fletcher11}.

To test the plausibility of the above discussed scenarios, sophisticated radiative transfer and dynamical calculations will be required that include heterogeneous clouds and consider the fast rotation of this object. 

We have measured longitudinal variability throughout the near-infrared in a T6 dwarf and found an unusual correlation of light curve phase with the pressure probed by a given wavelength, which suggests a complex horizontal and vertical atmospheric structure. Our observations should provide an incentive to drive the development of higher-dimensional atmospheric models in order to gain a deeper understanding of dynamical and radiative processes in brown dwarf and exoplanet atmospheres.

\acknowledgments

Support for Program number 12314 was provided by NASA through a grant from the Space Telescope Science Institute, which is operated by the Association of Universities for Research in Astronomy, Incorporated, under NASA contract NAS5-26555. This work is based on observations made with the Spitzer Space Telescope, which is operated by the Jet Propulsion Laboratory, California Institute of Technology under a contract with NASA. Support for this work was provided by NASA through award issued by JPL/Caltech (1439915).

\clearpage

\end{document}